\documentstyle[aps,twocolumn,epsf]{revtex}

\def\be{\begin{equation}}
\def\ee{\end{equation}}

\newcounter{fig}

\catcode`\@=11   

\newcommand{\fcaption}[1]{\vspace{1ex}   
        \refstepcounter{figure}   
        \setbox\@tempboxa = \hbox{\footnotesize {\bf Fig.~\thefigure.} #1}   
        \ifdim \wd\@tempboxa > 7cm   
           {\begin{center}   
        \parbox{7cm}{\footnotesize\baselineskip=8pt {\bf Fig.~\thefigure.} #1}   
            \end{center}}   
        \else   
             {\begin{center}   
             {\footnotesize {\bf Fig.~\thefigure.} #1}   
              \end{center}}   
        \fi}

\begin{document}

\title{Stokes formula and density perturbances 
for driven tracer diffusion in an adsorbed monolayer
}

\author{ O.B{\'e}nichou$^{1}$, A.M.Cazabat$^{2}$, J.De Coninck$^3$,
M.Moreau$^{1}$
 and G.Oshanin$^{1}$
}

\address{  $^{1}$ Laboratoire de Physique Th{\'e}orique des Liquides (CNRS - UMR 7600), 
Universit{\'e} Pierre et Marie Curie,
4 place Jussieu, 75252 Paris Cedex 05, France
}

\address{ $^{2}$ Laboratoire de Physique de la Mati{\`e}re Condens{\'e}e,
Coll{\`e}ge de France, 11 place M.Berthelot, 75231 Paris Cedex 05, France
} 

\address{ $^{3}$ Centre de Recherche en Mod\'elisation Mol\'eculaire,
Universit\'e de Mons-Hainaut, 20 Place du Parc, 7000 Mons, Belgium
}

\address{\rm (Received: )}
\address{\mbox{ }}
\address{\parbox{14cm}{\rm \mbox{ }\mbox{ }
We study the intrinsic friction of 
monolayers adsorbed 
on solid surfaces from
 a gas phase or vapor. Within the framework 
of the  Langmuir model of delocalized adsorption, we
calculate the resistance offered by 
the mobile adsorbate's particles to some 
impure tracer
molecule, 
whose diffusive random motion is biased by a constant external force.   
We find that for sufficiently small driving forces 
the force exerted on the 
tracer shows viscous-like behavior. 
We derive then the analog of the Stokes formula for 
two-dimensional adsorbates, 
calculate the corresponding friction coefficient 
and determine the stationary particle distribution in the monolayer 
as seen from the driven impurity.
}}
\address{\mbox{ }}
\address{\parbox{14cm}{\rm PACS No:  05.40-a,  66.30.Lw, 68.45.Da}}
\maketitle

\makeatletter
\global\@specialpagefalse

\makeatother

\pagebreak

When an ambient gas phase or vapor is brought in contact with
a solid, some portion of it becomes reversibly attached to the solid surface 
in form of an adsorbed layer.
 Such layers are important for various 
technological and material processing operations, including, for instance, 
coating, gluing or lubrication.

Following the work of Langmuir, 
equilibrium properties of
the adsorbates
have been extensively studied and a number of significant developments have been made
\cite{surf}. As well, 
some approximate results have been obtained for
both dynamics of an isolated adatom
on a corrugated surface and collective
diffusion, describing spreading of the macroscopic 
density fluctuations in
 interacting adsorbates being in contact with 
the vapor 
\cite{kreuzer,gortel}.

Another important aspect of dynamical behavior concerns tracer
 diffusion in adsorbates,
which is observed experimentally in  
STM or field ion measurements
 and provides
a useful information about adsorbate's viscosity or intrinsic friction. 
In this regard, analysis
 of tracer diffusion is not only a challenging question
in its own right, but is also
crucial for understanding of various dynamical processes taking
place on solid surfaces. To name but a few, we mention 
spreading of molecular 
films on solid surfaces \cite{spreading1}, 
spontaneous or forced dewetting of monolayers 
\cite{aussere}
or island formation \cite{islands}. However, 
apart of a slightly artificial 1D model \cite{olivier},
available studies of tracer diffusion in adsorbed layers 
focus on
strictly two-dimensional (2D) situations 
(see, e.g. Refs.\cite{kehr,elliott,deconinck}), excluding the
 possibility of
particles exchanges with the vapor.

In this Letter we present first results on intrinsic friction
in 
2D adsorbed monolayers composed of mobile particles
undergoing continuous exchanges
with the vapor phase. 
The system we consider here corresponds to
the generalized Langmuir model 
of adsorption with  
adsorbate lateral diffusion  \cite{surf} and
consists of three key ingredients: (i)
a solid surface containing some concentration of adsorption sites,  
(ii) a vapor phase and (iii)
a  monolayer of adsorbed hard-core particles, which  
perform activated random motion between the adsorption sites subject to 
the hard-core exclusion interactions 
(one particle per adsorption site at
most),
and undergo
continuous exchanges (desorption/adsorption) 
with the vapor phase (Fig.1).

\begin{figure} \begin{center}   
  \fbox{\epsfysize=6cm\epsfbox{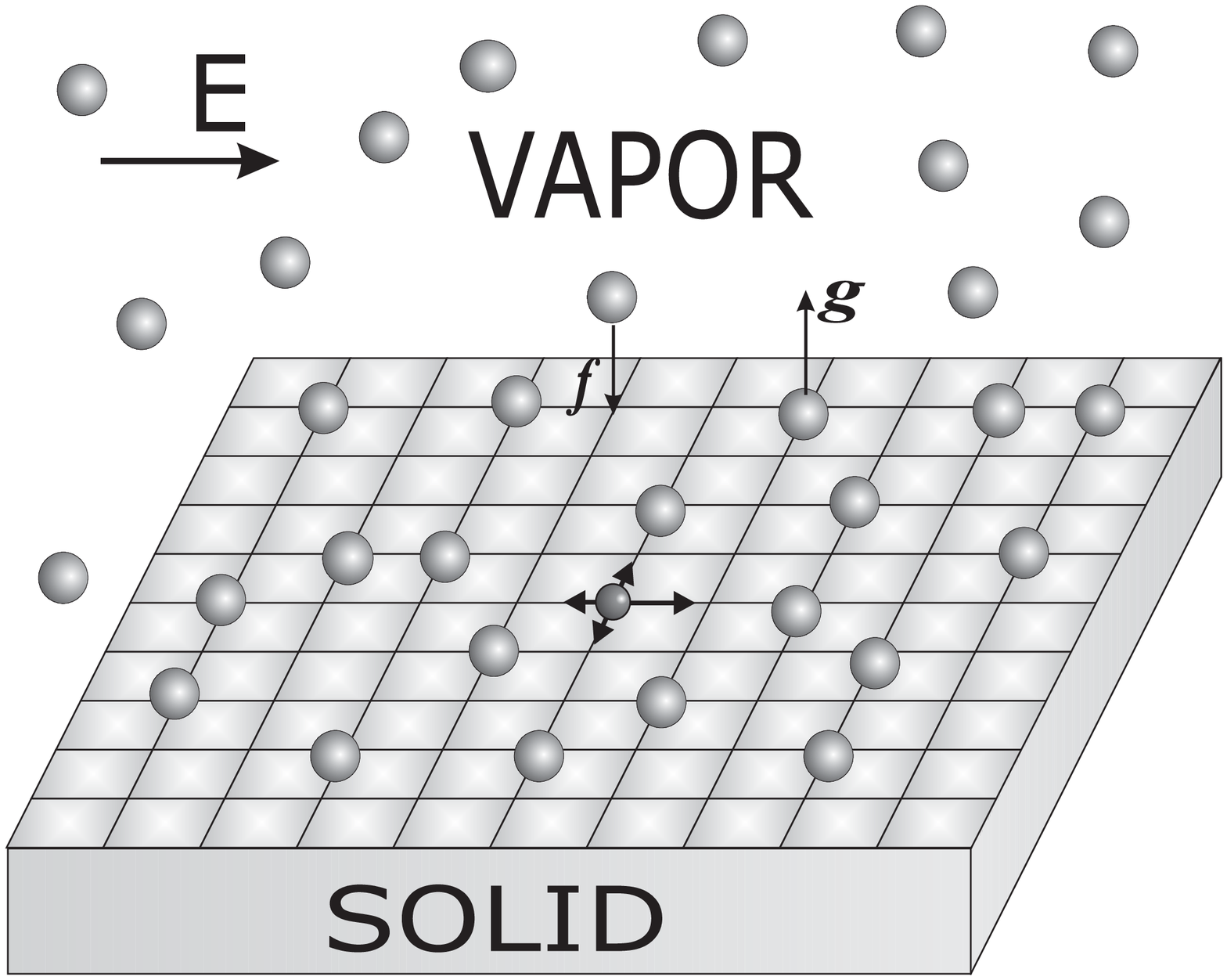}}   
   \fcaption{Adsorbed monolayer in contact with a vapor. 
Grey spheres denote the
monolayer (vapor) particles; the smaller black sphere 
stands for the driven tracer
particle.}   
  \label{tra}   
\end{center}\end{figure}

To determine the intrinsic friction of such an  
adsorbate we wish to probe
the resistance offered by the monolayer particles
 to some external
perturbance.
To this purpose, we add
an extra hard-core particle - 
a tracer, which may
move along the surface only and
can not desorb. Next, 
we suppose that the tracer is subject to an
external force $F$, 
which favors its motion in the preferential direction such that 
it ultimately attains a constant velocity $V$. 
Our aim is then to evaluate the velocity-force relation $V(F)$  
in the limit of small $F$
and to determine the
stationary particle distribution
in the perturbed monolayer.
General results for arbitrary $F$ will be
presented elsewhere.

More precisely, our model is defined as follows (Fig.1): 
We suppose first that the adsorption
sites 
form  a regular 2D square
lattice of spacing $\sigma$. 
Occupation of the 
lattice sites is characterized by
the set of time-dependent variables $\eta_{\vec{r}}$, 
$\vec{r}$ being the lattice vector 
of the site with Cartesian
coordinates $(x,y)$; 
$\eta_{\vec{r}}$ equals $1$ if the site $\vec{r}$ 
is occupied by an adsorbed particle and zero otherwise.

Next, we suppose that the particles from the vapor, (considered as a reservoir, 
which is maintained at a
constant pressure), 
 may  adsorb 
onto any vacant lattice site at a fixed rate $f/\tau^*$. 
Then, 
the adsorbed particles may move randomly by  
hopping with a rate $l/ 4 \tau^*$ to any of four
neighboring lattice sites,
which process is 
constrained by hard-core exclusion preventing multiple occupancy of
any site, 
or desorb from the lattice
at a site- and environment-independent rate $g/\tau^*$. 
For simplicity of exposition, we assumed here that  
typical adsorption, desorption and jump times
are equal to each other; these times, i.e. $\tau_{ad}$,
$\tau_{des}$ and 
$\tau_{jump}$, respectively, can be readily 
restored in our final results by the mere replacement
 $f \to \tau^{*} f/\tau_{ad }$, $g \to \tau^{*} g/\tau_{des}$ and
$l \to l \tau^*/\tau_{jump}$.

The tracer particle is initially put at the lattice origin.  
We stipulate
that this only particle is not allowed to
 desorb and in addition to random thermal forces experiences 
an action of a uniform external force $\vec{F} = (F,0)$. 
Dynamics of the tracer is then defined via standard rules \cite{lebowitz}: 
the tracer, 
which is at the site $\vec{R}$, $\vec{R} = (X,Y)$, 
 at time $t$,
 waits an exponentially distributed time 
with mean  $\tau$, (which is in the general case
 different from $\tau^*$),
 and then attempts to hop onto one of the 
neighboring sites, $\vec{R'}$, with probability
$p(\vec{R}|\vec{R'})$, 
defined as
\be
p(\vec{R}|\vec{R'}) = exp\Big(\frac{\beta}{2}(\vec{F}  
\vec{e}_{\nu})\Big) / \sum_{\mu}
exp\Big(\frac{\beta}{2}(\vec{F}  
\vec{e}_{\mu})\Big),
\ee 
where $\vec{e}_\nu = \vec{R'} - \vec{R}$, $\nu, \mu = 1,2, .., z$, $z$ being 
the coordination number of the lattice,  while 
$\beta$ is the reciprocal temperature. 
 The hop is fulfilled only if 
the target site appears to be
vacant; otherwise, the 
tracer remains at $\vec{R}$.

Now, several comments are in order.
First, 
we note that particles diffusion does not matter
in case when the tracer and the
inter-particle attractive interactions are absent; here, 
the monolayer is homogeneous and
the local occupation variables tend  as $t \to \infty$
to the same value $\rho_L = 1/(1+g/f)$, the latter 
relation being  the Langmuir adsorption
isotherm \cite{surf}. 
In the  presence of an impure molecule, whose hopping rates differ
from that of the monolayer particles, 
particles diffusion is significant and couples effectively
the occupation variables of different lattice sites.
This results, as we proceed to show, in the appearence of rather complex 
density
profiles. 
Second, we remark that in the model under study
the number of adsorbed particles is not explicitly conserved. 
The conserved particles number 
(CPN) limit can be achieved in the
stationary regime
by setting $f$ and $g$ to zero, while keeping 
their ratio fixed, $f/g = \rho_L/(1-\rho_L)$. 
This limit corresponds to 
standard models 
of tracer diffusion in 2D hard-core lattice gases.  Lastly, we remark that our model
can be thought off as a certain generalization of a model for dynamic percolation
proposed in \cite{grest}.

Let $P(\vec{R},\eta)$ denote the probability of finding  
at time $t$ the tracer at the site $\vec{R}$
and all adsorbed particles in the configuration $\eta = \{\eta_{\vec{r}}\}$. 
Further more, let $\eta^{\vec{r},\nu}$ denote the 
configuration obtained from $\eta$ by the Kawasaki-type 
exchange of the occupation
variables of the two neighboring 
sites $\vec{r}$ and $\vec{r} + \vec{e}_\nu$,
 and 
$\tilde{\eta}^{\vec{r}}$ be the configuration obtained from 
$\eta$ by
the replacement $\eta_{\vec{r}} \to 1-\eta_{\vec{r}}$,
 which corresponds to the
Glauber-type flip of the occupation variable due to 
the adsorption/desorption 
events. 
Then, counting up all events which can result in the
 configuration 
$(\vec{R},\eta)$ at time $t$ or modify it, 
we write down the following
master equation

\begin{eqnarray}
& &\partial_t P(\vec{R},\eta)= \frac{1}{\tau} \sum_{\nu} \Big[ p(0|\vec{e}_{\nu}) P(\vec{R}
 - \vec{e}_\nu,\eta) ( 1 - \eta_{\vec{R}}) - \nonumber\\
&-& p(0|\vec{e}_\nu) P(\vec{R},\eta) ( 1 - \eta_{\vec{R} +\vec{e}_\nu})\Big] + \frac{l}{4\tau^*}
\sum_{\vec{r},\nu} \; \Big[P(\vec{R},\eta^{\vec{r},\nu}) - \nonumber\\
&-& P(\vec{R},\eta)\Big] + \frac{g}{\tau^*} \sum_{\vec{r}} \Big[(1-\eta_{\vec{r}})
 P(\vec{R},\tilde{\eta}^{\vec{r}})-\eta_{\vec{r}} P(\vec{R},\eta)\Big] +\nonumber\\
& &+\frac{f}{\tau^*} \sum_{\vec{r}} \; \Big[\eta_{\vec{r}} P(\vec{R},\tilde{\eta}^{\vec{r}}) 
-(1-\eta_{\vec{r}}) P(\vec{R},\eta)\Big],
\label{eq maitresse}
\end{eqnarray} 
where the symbol $\vec{r}$
under the summation sign means that the sum
runs over all the lattice sites.

Now, we are in position to obtain the $X$-component of the 
tracer velocity $V$.
Multiplying both sides of Eq.(\ref{eq maitresse}) by
$X$ and summing over all $(\vec{R},\eta)$, we have
\begin{eqnarray}
& &V_X=
\frac{\sigma}{\tau}\Big[ p(0|(\sigma,0))\Big(1-k(\sigma,0)\Big)- 
 p(0|(-\sigma,0)) \times\nonumber\\
& &\times
\Big(1-k(-\sigma,0)\Big)\Big], \; \; \; k(\vec{\lambda})=
\sum_{\vec{R},\eta}\eta_{\vec{R}+\vec{\lambda}},
P(\vec{R},\eta)
\label{vitesse}
\end{eqnarray}
where $k(\vec{\lambda})$
is the probability of having at time t an adsorbed particle 
at position $\vec{\lambda}=(\lambda_x,\lambda_y)$, defined 
in the frame of reference moving with the
tracer particle. In other words, $k(\vec{\lambda})$
 can be interpreted 
as the density
profile as seen from the moving tracer.

Consequently, $V_X$ depends on the particle density in the 
immediate vicinity of the tracer. Note, 
that if the monolayer is perfectly stirred, i.e. 
$k(\vec{\lambda}) = \rho_L$, (which implies
 decoupling of the occupation variables),  we would obtain
from Eq.(\ref{vitesse})  a trivial mean-field result
$V_X = V_{0} = \sigma (p-q)  (1 - \rho_L)/\tau$,
which states that the tracer jump time $\tau$ gets
 merely renormalized by a
 factor $(1 - \rho_L)^{-1}$, which  defines 
the frequency of $\em successful$ jump events. 
However, this
is not the case and  $k(\vec{\lambda}) \neq \rho_L$,
 except for $|\lambda| \to \infty$. 
 Moreover,   
$k(\vec{\lambda})$ is a complicated function of $V_X$,
 which renders Eq.(\ref{vitesse})
to be strongly non-linear. 

Equations describing
the time evolution of $k(\vec{\lambda})$ can be found from
(\ref{eq maitresse}) by multiplying both sides of it by 
$\eta_{\vec{R} + \vec{\lambda}}$  
and summing over all
configurations $(\vec{R}, \eta)$. In doing so, we find 
that these equations are not closed with respect to 
$k(\vec{\lambda})$, 
but are coupled to the third-order
correlations,
\begin{equation} 
\label{T}
T(\vec{\lambda},\vec{e}_\nu)
 = \sum_{\vec{R}, \eta} \eta_{\vec{R} + \vec{\lambda}} \eta_{\vec{R} +
\vec{e}_\nu} P(\vec{R},\eta)
\end{equation}
In turn, if we proceed further to
the third-order
correlations, we find that these are 
coupled respectively to the fourth-order correlations. 
Consequently, in order
to compute $V_X$, one faces
the problem of solving an infinite hierarchy of coupled equations
for the correlation functions.  
Here we resort to the simplest non-trivial 
closure of the hierarchy    
in terms of $k(\vec{\lambda})$ representing $T(\vec{\lambda},\vec{e}_\nu)$ 
as

\be
T(\vec{\lambda},\vec{e}_\nu) = \sum_{\vec{R}, \eta} \eta_{\vec{R} + \vec{\lambda}} 
P(\vec{R},\eta) \sum_{\vec{R}, \eta} \eta_{\vec{e}_\nu} P(\vec{R},\eta)
\label{decouplage}
\ee

We hasten to remark that the decoupling 
in Eq.(\ref{decouplage}) provides exact 
results for the analogous 1D model in the CPN limit
\cite{burlatsky,olla} and serves as a very good approximation 
for the 1D model with a reservoir \cite{olivier}.
We set out to show in what follows that 
in the CPN limit
our results reproduce exactly the classic results
of Nakazato and Kitahara \cite{elliott}, which are exact for
$\rho_L \ll 1$ or $\rho_L \sim 1$, and serve as a very 
good approximation for any intermediate
$\rho_L$ \cite{kehr}. Since adsorption/desorption processes
 are essentially linear, we expect that
such a closure will provide an accurate description for 
arbitrary $f$ and
$g$.

Employing the approximation in Eq.(\ref{decouplage}),
 we obtain for $k(\vec{\lambda})$ 
the following closed evolution equations
\begin{eqnarray}
\label{eq approx}
& &\partial_t k(\vec{\lambda})={\sl \tilde{L}} \; k(\vec{\lambda})
+\frac{f}{\tau^*}, \; \; \; {\sl \tilde{L}} \equiv \frac{l}{4\tau^*} \triangle - \nonumber\\
& &-\frac{f+g}{\tau^*} +\frac{1}{\tau} \sum_{\nu} p(0|\vec{e}_\nu) \Big(1 - k(\vec{e}_\nu)\Big)
\nabla_{\vec{e}_\nu},
\end{eqnarray}
\be
\triangle k(\vec{\lambda}) 
\equiv \sum_{\nu} \nabla_{\vec{e}_\nu}  k(\vec{\lambda}) \equiv 
\sum_{\nu} \Big[ k(\vec{\lambda} + \vec{e}_{\nu})  -  k(\vec{\lambda})\Big]
\ee

Equation (\ref{eq approx}) holds for any site
 $\vec{\lambda}$, except for the four sites in the
immediate vicinity of the tracer:
$k(\vec{\lambda})$ at these four
neighboring sites is essentially perturbed 
due to the asymmetric hopping rules of the tracer 
and obeys
\begin{eqnarray}
& &\dot{k}(\vec{e}_\nu)= \left . \Big({\sl \tilde{L}} \; k(\vec{\lambda})\Big)
\right|_{\vec{\lambda}=\vec{e}_\nu} + \frac{1}{\tau} p(0|\vec{e}_\nu)
 k(2 \vec{e}_\nu) \Big(1-k(\vec{e}_\nu)\Big) - \nonumber\\
&-&\frac{1}{\tau} k(\vec{e}_\nu) p(\vec{e}_\nu|0) \Big(1-k(-\vec{e}_\nu)\Big)   
-\frac{l}{4\tau^*}\nabla_{-\vec{e}_\nu} k(0) + \frac{f}{\tau^*}
\label{limite1}
\end{eqnarray}
Note that Eqs(\ref{eq approx}) to
(\ref{limite1}) constitute a closed system of
equations, which suffice in principle the
 computation of the density profiles
and of the tracer velocity. However, these equations are
non-linear, since $k(\vec{e}_\nu)$  
enters the prefactor before the gradient terms,
which makes
 such a
computation to be a  non-trivial problem.  
Below we consider the stationary solution of Eqs.(\ref{eq approx}) and
(\ref{limite1}).

Solution of Eqs.(\ref{eq approx}) to  (\ref{limite1}) in the limit $t \to \infty$ 
can be readily obtained by applying the generating function approach and reads:
\begin{eqnarray}
& & k(\vec{\lambda}) = \rho_L + \Big(\frac{\sigma}{2 \pi \sqrt{\alpha_F}} \Big)^2  \Big[ \sum_{\nu}
A_{\vec{e}_\nu}  \Big(k(\vec{e}_\nu) - 
 \rho_L\Big) \nabla_{-\vec{e}_\nu} {\cal F}_{\vec{\lambda}} - \nonumber\\
& &- \rho_L \Big(A_{\sigma,0}  - A_{-\sigma,0} \Big) 
\Big(\nabla_{\vec{e}_\nu} {\cal F}_{\vec{\lambda}} - \nabla_{-\vec{e}_\nu}
{\cal F}_{\vec{\lambda}}\Big)\Big],
\label{profs}
\end{eqnarray}
where $A_{\vec{e}_\nu}  = 1 + 4 \tau^*
 p(0 | \vec{e}_\nu) (1 - k(\vec{e}_\nu))/l \tau$, 
$\alpha_F = \sum_{\nu} A_{\vec{e}_\nu}  +  4 g/l(1 - \rho_{L})$, 
$\theta = 2
(\sqrt{A_{\sigma,0}  A_{-\sigma,0} } + A_{0,\sigma} )/\alpha_F$, while
\begin{equation}
\label{calculdeF}
{\cal F}_{\vec{\lambda}}= 
 \left(\frac{A_{-\sigma,0} }{A_{\sigma,0} }\right)^{\lambda_x/2 \sigma}
\int^{\pi}_{-\pi} \int^{\pi}_{-\pi} d^2 \vec{k}
\frac{exp\Big( - i (\vec{k } \vec{\lambda }) \Big)}{1 - \theta \; \xi(\vec{k})},
\end{equation}
and
\begin{equation}
\label{struc}
\xi(\vec{k}) = \frac{ \sqrt{A_{\sigma,0}  A_{-\sigma,0} } \;
\cos(\sigma k_x)+A_{0,\sigma}  \;\cos(\sigma k_y)}{\sqrt{A_{\sigma,0} 
 A_{-\sigma,0} }+A_{0,\sigma} }
\end{equation}
Lastly,  we have to determine four particular values
$k(\vec{e}_\nu)$ which appear on the rhs of Eq.(\ref{profs}). This can
 be done by setting
 $\vec{\lambda} = \vec{e}_\nu$, $\nu = 1,2, .., z$,  to the lhs of  Eq.(\ref{profs}) and 
solving the resulting system of four
linear equations. In doing so, we define $k(\vec{e}_\nu)$ explicitly, as 
functions of $V_X$ and of the characteristic microscopic parameters.

Consider now the behavior of the tracer velocity $V_{X}$ in the
 limit $\sigma \beta F \ll 1$.  
First,  we have from
Eq.(\ref{vitesse}) that 
\begin{equation}
V_X \sim \frac{\sigma}{4 \tau}\left\{\beta\sigma F(1-\rho_L)-\delta k \right\},
\label{eq:14}
\end{equation}
where $\delta k \equiv 
k(\sigma,0)-k(-\sigma,0)$ denotes the density jump in the vicinity
 of the tracer. Note, that since $\delta k > 0$, the tracer velocity $V_X$
is always smaller than the mean-field prediction $V_0$. Next, 
from Eqs.(\ref{profs}) to (\ref{struc}) we find 
\begin{eqnarray} \label{tri}
& &\frac{\delta k}{\sigma \beta F} 
\sim 2 \rho_L (1-\rho_L)/ 
 \Big[ 3 \rho_L - 1 - \frac{l \tau}{\tau^*} + \frac{l \tau A_{\sigma,0}(F = 0)}{\tau^*}/  \nonumber\\
& &\int_0^\infty \frac{dw}{w} \;  
\exp\Big(-\frac{\alpha_0 \; w}{2 A_{\sigma,0}(F = 0)}\Big) \; I_1(w) \; I_0(w)\Big],
\end{eqnarray}
$I_n(w)$ being the modified Bessel function of order $n$.

Consequently, we find from Eq.(\ref{eq:14}) that in the
 limit $\sigma \beta F \ll 1$ the
force-velocity relation attains the physically revealing
 form of the Stokes formula,
i.e. $ F = \zeta V_X$, which signifies that the friction
 force exerted by the
monolayer particles on the driven tracer is viscous. 
The corresponding friction coefficient
$\zeta$ is given by
\be
\label{friction}
\zeta = \zeta_0 + \zeta_{coop},
\ee
where $\zeta_0 = 4 \tau/\sigma^2 \beta (1 - \rho_L)$ and $\zeta_{coop}$ follows
\be
\zeta_{coop} = \zeta_0 \frac{(\delta k/\sigma \beta F)}{1 - \rho_L - (\delta k/\sigma \beta F)},
\ee
$(\delta k/\sigma \beta F)$ being defined in Eq.(\ref{tri}). 
 While $\zeta_0$
is the typical mean-field-type result
 (see the discussion following Eq.(\ref{vitesse})),
 the second term has a more complicated origin and
is associated with the cooperative behavior: 
de-homogenization of the monolayer by driven impure molecule 
and formation 
 of stationary density profiles, whose characteristic parameters 
depend on the velocity $V_X$. 
This second term can be small when either 
$\rho_L \ll 1$, or $\tau^* \to 0$ (perfect mixing), but 
 dominates the overall friction for any intermediate values of
 systems parameters. 
We finally remark that in the CPN limit
Eq.(\ref{friction}) reduces to the classical result obtained in \cite{elliott}, 
which is known to be exact when $\rho_L \ll 1$ or $\rho_L \sim 1$
and represents a very good approximation for 
any intermediate value of $\rho_L$ \cite{kehr}.

Lastly, we discuss the characteristic features of the monolayer 
density profiles as seen from the stationary moving tracer,  Eq.(\ref{profs}),
at large distances in front of and past the tracer. 
It follows from Eq.(\ref{profs}) that in the limit $|\lambda_x| \to \infty$, 
($\lambda_y = 0$)
the density profiles obey
\be
\label{cor}
k(\lambda_x,0) \sim \rho_L \pm \frac{\triangle}{2} \Big(\frac{\sigma}{|\lambda_x|}\Big)^{1/2}  exp\Big( \pm  
\frac{|\lambda_x|}{\sigma} ln(z_{\pm}) \Big),
\ee
where the sign "+" ("-") corresponds to the domain $\lambda_x > 0$ ($\lambda_x < 0$), 
while $z_{\pm}$ are two
of four eigenvalues of
the operator ${\sl \tilde{L}}$ in Eq.(\ref{eq approx}), which are given by
\be
z_{\pm} = A^{-1}_{-\sigma,0} \Big(\frac{\alpha_{F}}{2} - A_{0,\sigma} \Big) \Big[1 
\pm \sqrt{1 - \frac{2 A_{\sigma,0} A_{-\sigma,0}}{\alpha_{F} - 2 A_{0,\sigma}}}\Big]
\ee
Hence, in the domain $\lambda_x > 0$ the monolayer density is 
higher than $\rho_{L}$, which means 
that there is a "traffic jam" type region in 
front of the tracer, which impedes its motion.  This region decays
exponentially with the distance. In turn, past the 
tracer there is an exponentially decaying depleted region in which the 
density is lower than $\rho_{L}$; since $z_- < z_+$, the
depleted region is more extended in space than the
 condensed one in front of the tracer, such that the density profiles 
are asymmetric with respect to $\lambda_x = 0$.
It is interesting to note that for $k(\vec{\lambda})$ as in Eq.(\ref{profs}) the sum
 $\sum_{\vec{\lambda}} (k(\vec{\lambda}) - \rho_L) \equiv 0$, which
 means that the tracer does not
perturb the global balance
between adsorption and desorption. This is not, however, an $\em a
\; priori$ evident result in view of the asymmetry of the density profiles.

The salient feature of the behavior past the tracer is that in the CPN limit
$z_- \equiv 1$ 
 at any fixed $F$, which
signals that Eq.(\ref{cor})
 is no longer valid and correlations get somewhat stronger.
Indeed, we find that
\be
\label{power}
k(\lambda_x,0) - \rho_L \sim - const \times |\lambda_{x}|^{-3/2}, \; \; \lambda_x \to
-\infty
\ee
i.e. the correlations 
fall off with the distance as a 
$power-law$! Remarkably, this implies that in the CPN limit
the mixing of the monolayer 
is not at all efficient and there are considerable memory effects - 
the host medium  remembers
the passage of the tracer on a large space and time scales. This situation
can be realized experimentally for ultrathin 
liquid films confined in narrow
slits between solid surfaces, e.g., in boundary lubrication. We note
also that the power-law behavior in Eq.(\ref{power}) can be observed as an
intermediate scale decay for adsorbed layers exposed to a low vapor pressure.

To conclude,  we have studied analytically the
 intrinsic frictional properties
 of 2D adsorbed monolayers,
composed of mobile hard-core particles  
undergoing continuous exchanged with the vapor. By analysing the
 force-velocity relation of a driven impure molecule - a tracer, 
which is designed to  probe the resistance offered
 by adsorbate particles to the external perturbance, we have
derived the analog of the Stokes formula for 2D mobile 
adsorbates  and calculated explicitly the corresponding friction 
coefficient. Besides, we have determined the stationary
density profiles, which emerge in the adsorbate in response to the presence of a 
driven impurity,
as well as obtained explicit results for both the density jump in the
immediate vicinity of the tracer and  
the asymptotical density relaxation at large separations from it.

\end{document}